\documentclass[twocolumn,showpacs,preprintnumbers,amsmath,amssymb]{revtex4}

\usepackage{graphicx}
\usepackage{dcolumn}
\usepackage{bm}

\begin{document}


\title{Hydrogen Dissociation and Diffusion on Ni and Ti -doped
Mg(0001) Surfaces}

\author{M. Pozzo$^{1,3}$}
\author{D. Alf\`{e}$^{1,2,3,4}$}%
\email{d.alfe@ucl.ac.uk} 
\affiliation{
$^1$Department of Earth Sciences, University College London, 
Gower Street, London WC1E 6BT, United Kingdom \\
$^2$Department of Physics and Astronomy, University College London, Gower
Street, London WC1E 6BT, United Kingdom \\
$^3$Material Simulation Laboratory, University College London, 
Gower Street, London WC1E 6BT, United Kingdom \\
$^4$London Centre for Nanotechnology, University College London, 17-19 Gordon
Street, London WC1H 0AH, United Kingdom}%

\author{A. Amieiro, S. French and A. Pratt}
 \homepage{http://www.matthey.com}
\affiliation{Johnson Matthey Plc, Technology Centre, Blounts Court, Sonning Common, Reading, 
RG4 9NH, United Kingdom \\} 
\date{\today}

\begin{abstract}
It is well known, both theoretically and experimentally, that alloying
MgH$_2$ with transition elements can significantly improve the
thermodynamic and kinetic properties for H$_2$ desorption, as well as
the H$_2$ intake by Mg bulk.  Here we present a density functional
theory investigation of hydrogen dissociation and surface diffusion
over Ni-doped surface, and compare the findings to previously
investigated Ti-doped Mg(0001) and pure Mg(0001) surfaces. Our results
show that the energy barrier for hydrogen dissociation on the pure
Mg(0001) surface is high, while it is small/null when Ni/Ti are added
to the surface as dopants. We find that the binding energy of the two
H atoms near the dissociation site is high on Ti, effectively impeding
diffusion away from the Ti site. By contrast, we find that on Ni the
energy barrier for diffusion is much reduced.  Therefore, although
both Ti and Ni promote H$_2$ dissociation, only Ni appears to be a
good catalyst for Mg hydrogenation, allowing diffusion away from the
catalytic sites.  Experimental results corroborate these theoretical
findings, i.e. faster hydrogenation of the Ni doped Mg sample as
opposed to the reference Mg or Ti doped Mg.

\end{abstract}

\maketitle

\section{Introduction}\label{intro}

Safe and efficient hydrogen storage is one of the biggest barriers to
the more wide spread usage of hydrogen as an energy carrier or fuel.
Currently, commercial solutions are based on liquid or compressed gas
storage methods, which are inefficient and have safety
issues. Alternative storage methods include metal hydrides, which are
formed by the interaction between a suitable metal and hydrogen. The
relatively strong metal-hydrogen bonds provide an intrinsically safe
storage medium. The release of hydrogen from the hydride is then
achieved by heating the material above a certain decomposition
temperature. There are a large number of metals in nature that form
hydrides, however, only the lighter ones are thought to be suitable
candidates for mobile hydrogen storage purposes (see
\cite{schlapzutt01} for an overview).  Beside being light weight, a
hydride will need to have good cyclability (several hundred times with
little loss of performance), fast adsorption/desorption kinetics (the
hydride should form/decompose on a time scale of minutes) and low
decomposition temperature (ideally between 20 and 100 Celsius).

Magnesium is a good case study due to its lightweight, low cost,
cyclability and the high H storage capacity of 7.6\% by weight once
the hydride MgH$_2$ is formed~\cite{schwarz99}. However, its
commercial application is still on hold for practical issues due to
low H absorption/desorption kinetics and high working
temperatures~\cite{bobet00}.  The strong bond between Mg and hydrogen
provides MgH$_2$ with high thermodynamic stability, which has an
enthalpy of formation of about -76 kJ/mol~\cite{yamaguchi94}, and a
decomposition temperature of more than 300
Celsius~\cite{bogdanovic99}.  The slow kinetics may be explained by
the high energy barrier which needs to be overcome (see for
example~\cite{sprunplu91}) to dissociate the H$_2$ molecule due to the
tendency of Mg to repel the s-electron of H because of the Pauli
exclusion principle~\cite{hamnor95}.

A step forward in improving hydrogen reaction kinetics has been
achieved by the mechanical ball milling of MgH$_2$ with transition
elements (see~\cite{oelerich01} and references therein). The hydrogen
storage properties of mechanically milled powders improves because of
the reduced powder size (see for example
\cite{zaluski95,liang99,zaluska99,shang04} and references therein),
which shorten the diffusion distance of H into bulk Mg for the
formation of the hydride. There are many experimental and theoretical
papers in the literature showing that the hydriding properties of
MgH$_2$ are further enhanced by the addition of traces of transition
metals which act as a catalyst (see for
example~\cite{yavari03,liang04,shang04,song04,vegge05} and references
therein).  In particular, alloying Mg with Ni can slightly improve the
thermodynamic properties of the hydride by favouring H$_2$
adsorption/dissociation and consequent atomic hydrogen
absorption/desorption due to a weakening of the bonding between Mg and
H atoms (see for example
~\cite{liang99,shang04,song04,vegge05,hanada05} and references
therein).

While from theoretical calculations the destabilization effect of Ni
on MgH$_2$ appears second only to Cu (see~\cite{song04}),
experimentally Ni shows the highest kinetics, with Cu falling
behind. As suggested by Shang et al.~\cite{shang04}, Cu results are
disadvantageous for H desorption probably because of the formation of
a MgCu$_2$ compound. Recently, a new method of chemical fluid
deposition in supercritical fluids has been used on metal
hydrides~\cite{bobet07}.  Even sparser literature exists for the
activation barrier of hydrogen dissociation on a transition metal
doped surface, which includes only the theoretical calculations made
by Du et al.~\cite{du05,du06,du07} within DFT (RPBE) for both the pure
Mg(0001), and Ti and Pd incorporated Mg surfaces. Their results show
that the dissociation barrier of hydrogen on the Ti doped Mg surface
is greatly reduced (in fact, there is no barrier at all) due to the
strong interaction between the hydrogen $s$ orbital and the Ti $d$
orbital, however, strong binding of the two H atoms near the Ti site
prevents easy diffusion, reducing therefore the efficacy of the
catalyst for Mg hydrogenation~\cite{du07}. Palladium doping appears to
both lower the dissociation barrier and the diffusion barrier,
suggesting a better catalytic activity.  Their findings are consistent
with the experimentally observed trend of generally improved hydrogen
absorption kinetics when Mg surfaces are doped with transition metals,
as previously mentioned.

To our knowledge, so far there is no published theoretical
investigation of H$_2$ dissociation and corresponding activation
barrier on a Ni-incorporated Mg surface, nor a systematic
investigation of the catalytic effect of other transition metal
dopants apart from the above mentioned studies on the Ti-doped and
Pd-doped Mg surfaces presented by Du et
al.~\cite{du05,du06,du07}. There are a few theoretical papers about
the dissociation of molecular hydrogen on a pure Mg surface where the
corresponding activation barrier has been effectively
calculated. These investigations were based on a jellium model and
potential energy surface(PES) calculations within density functional
theory (DFT) with the local density approximation(LDA) or generalised
gradient corrections
(RPBE)~\cite{hjelmberg79,johansson81,norskhoum81,bird93,vegge04}.

For the purpose of a larger scale investigation, we have performed DFT
calculations for hydrogen dissociation and diffusion on a Ni-doped Mg
surface, accompanied by analogous calculations on a Ti-doped Mg
surface for a consistency check with the recently reported theoretical
values. This study should be regarded as a first step in order to
build up a global picture of the dissociative chemisorption of
hydrogen when doping the Mg surface with different transition
metals. The main purpose of this article is to try to understand the
observed large enhancement of the kinetics of hydrogen adsorption by
Mg when it is doped with a small quantity of Ni, but not when it is
doped with Ti.  The computational results are supported by
experimental data where a Ni doped Mg sample is hydrogenated
substantially faster than the reference Mg or Ti doped Mg.

\section{Computational method}\label{methods}

\begin{figure}
\rotatebox{0}{\scalebox{0.6}[0.45]{\includegraphics[width=\columnwidth]{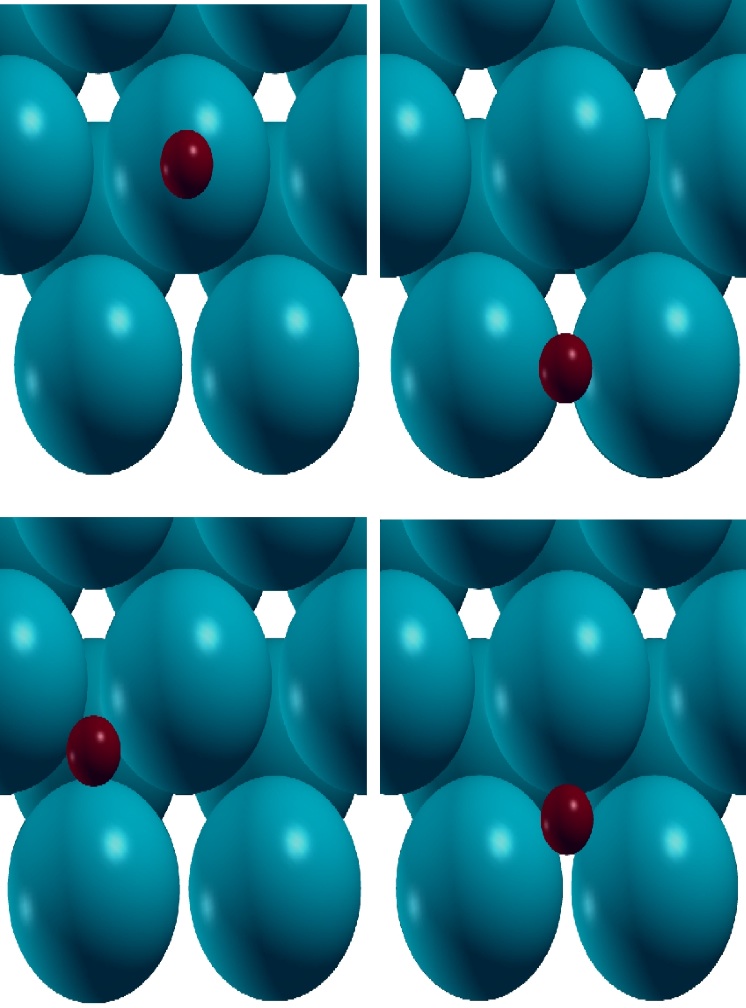}}}
\caption{\label{adssites} (Colour) Possible adsorption sites (top, bridge, hollow-hcp 
and hollow-fcc) for hydrogen (dark red) on the Mg(0001) surface (light blue).}
\end{figure}

\begin{figure}
\rotatebox{-90}{\scalebox{0.8}[0.9]{\includegraphics[width=\columnwidth]{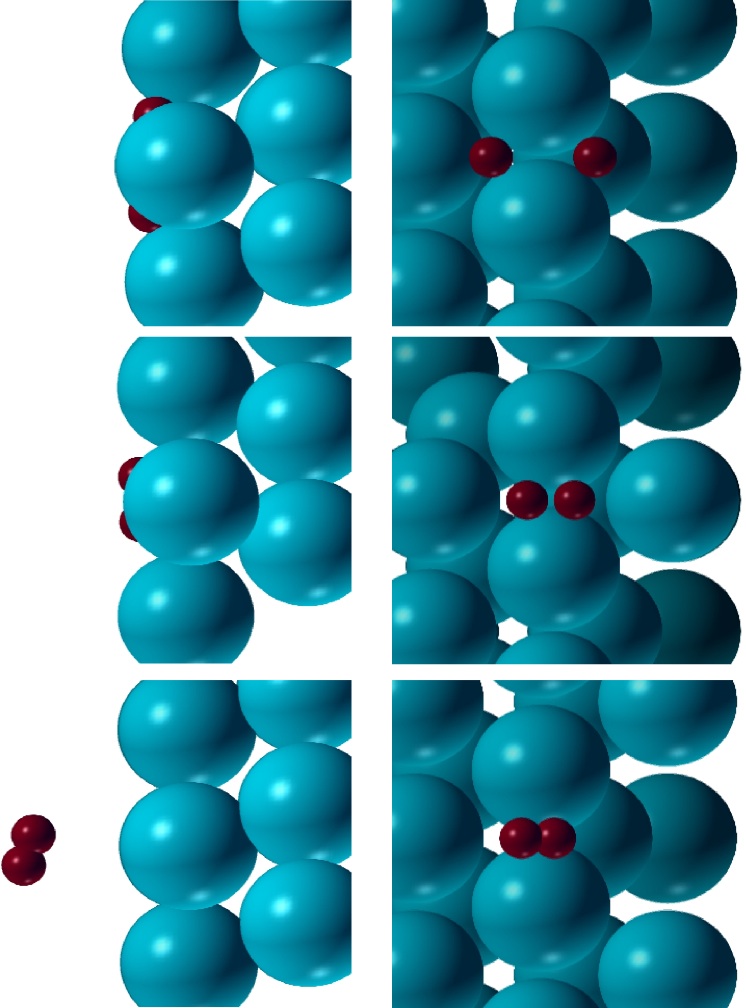}}}
\caption{\label{MgH2diss} (Colour) H$_2$ (dark red) dissociation on the pure Mg 
(light blue) surface as viewed from side (top figures) and top (bottom figures). 
Figures show positions at IS (left-hand panel), TS (central panel) and FS (right-hand panel).}
\end{figure}

\begin{figure}
\rotatebox{0}{\scalebox{0.9}[0.9]{\includegraphics[width=\columnwidth]{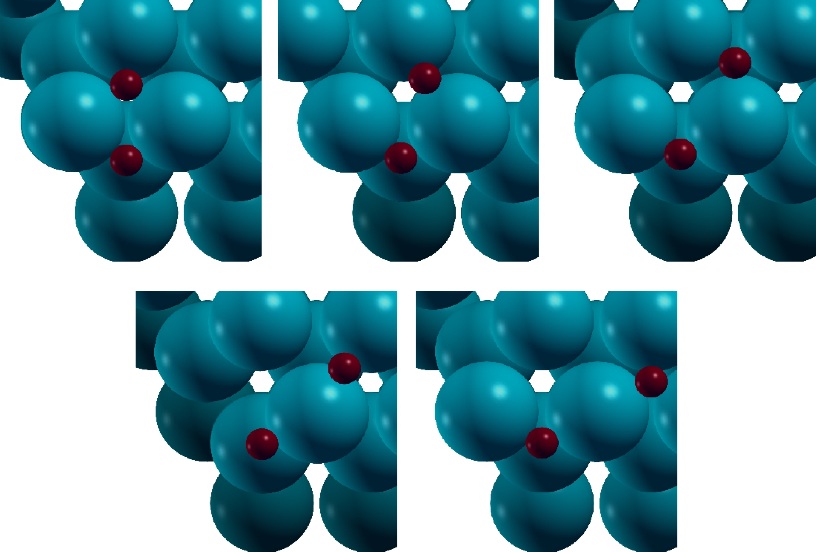}}}
\caption{\label{MgH2diff} (Colour) H (dark red) diffusion on the pure Mg (light blue) 
surface as viewed from top. Figures show positions at FS (top-left), TS2 (top-centre),
LS (top-right), TS3 (bottom-left) and FS2 (bottom-right).}
\end{figure}

\begin{figure}
\rotatebox{0}{\scalebox{0.6}[0.45]{\includegraphics[width=\columnwidth]{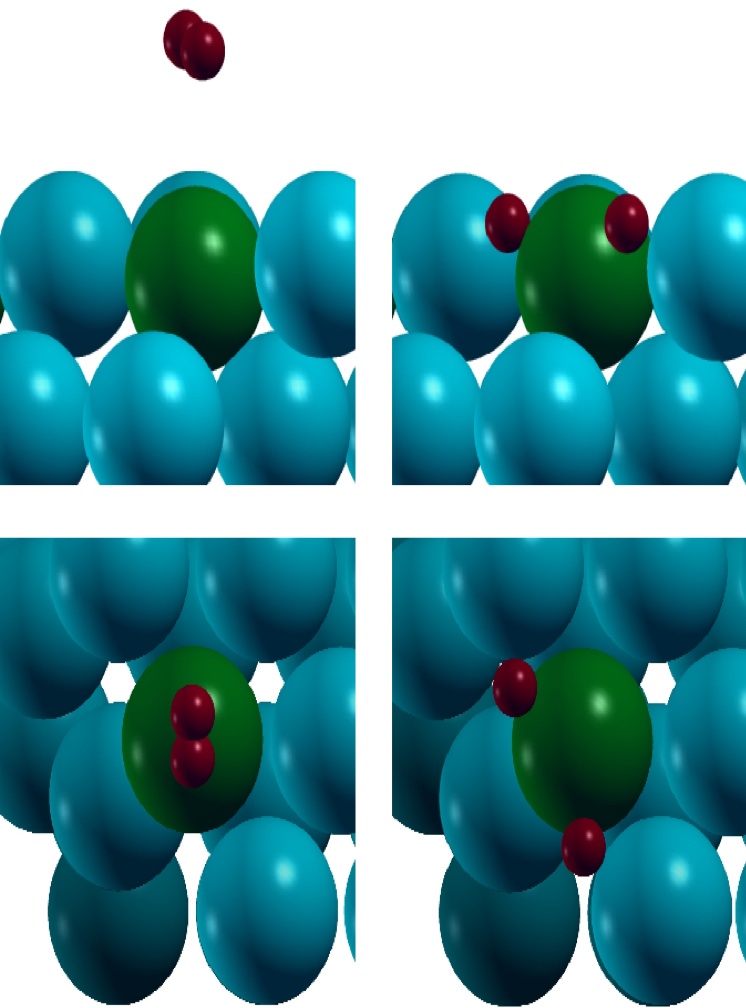}}}
\caption{\label{MgTiH2diss} (Colour) Same as Fig.~\ref{MgH2diss} but for H$_2$ 
dissociating over the Ti-doped Mg surface at IS and FS (there is no TS in this case).
The Mg, Ti and H atoms are represented respectively by light blue, dark green and 
dark red  colours.}
\end{figure}

\begin{figure}
\rotatebox{0}{\scalebox{0.95}[0.45]{\includegraphics[width=\columnwidth]{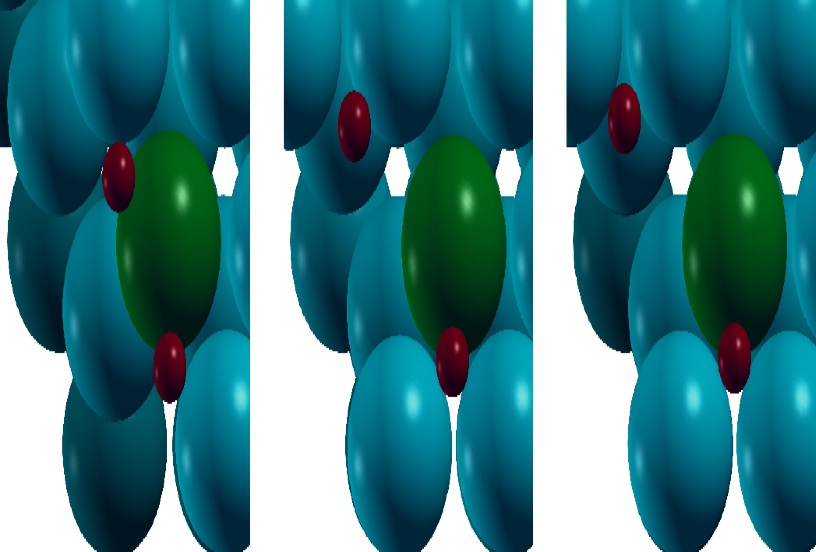}}}
\caption{\label{MgTiH2diff} (Colour) Same as Fig.~\ref{MgH2diff} but for H diffusion
over the Ti-doped Mg surface. Figures show positions at FS (left), TS2 (centre) and 
FS2 (right). The Mg, Ti and H atoms are represented respectively by light blue, dark green 
and dark red colours.}
\end{figure}

\begin{figure}
\rotatebox{0}{\scalebox{1.12}[0.42]{\includegraphics[width=\columnwidth]{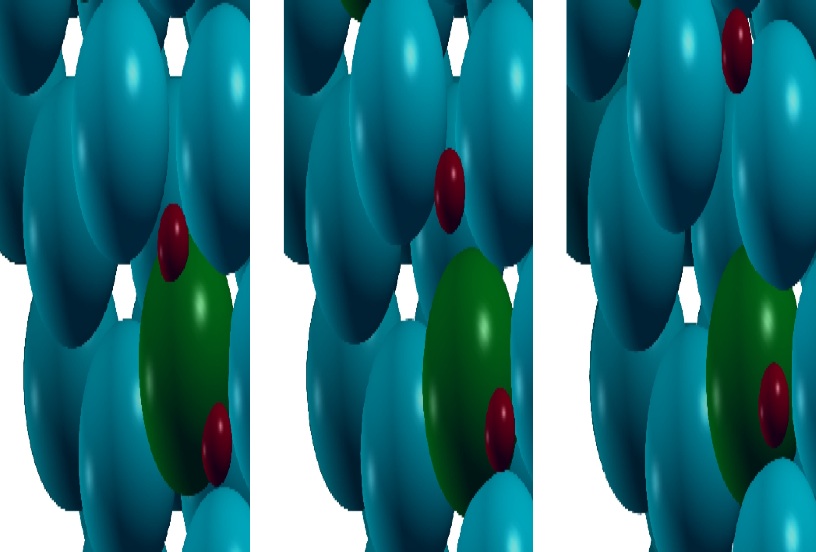}}}
\caption{\label{MgNiH2diff} (Colour) Same as Fig.~\ref{MgH2diff} but for H diffusion
over the Ni-doped Mg surface. Figures show positions at FS (left), TS2 (centre) and 
FS2 (right). The Mg, Ni and H atoms are represented respectively by light blue, dark green 
and dark red colours.}
\end{figure}

DFT calculations were performed with the ab-initio simulation package
VASP~\cite{vasp} using the projector augmented wave (PAW)
method~\cite{blochl94,kresse99} and the PBE exchange-correlation
functional~\cite{pbe}. An efficient charge density extrapolation was
used to speed up the calculations~\cite{alfe99}. A plane-wave basis
set was used to expand the electronic wave-functions with a plane-wave
energy cut-off of 270 eV, which guarantees convergence of adsorption
energies within 1 meV. For completeness, Mg bulk parameters were also
calculated using the LDA functional.  Monkhorst-Pack {\bf k}-points
were used to sample the Brillouin zone~\cite{monk}.  A smearing
function of the Methfessel-Paxton (MP) type (product of a Gaussian
times a $n$th-order Hermite polynomial)~\cite{methfessel89} was used
throughout.  Figs.~\ref{adssites}, \ref{MgH2diss} - \ref{MgNiH2diff} 
and \ref{charge} have been made using the XCRYSDEN 
software~\cite{xcrysden}. The exact values of the various parameters
used in the calculations will be reported below in the relevant
sections.

Activation energies have been calculated using the nudged elastic band
(NEB) method~\cite{neb}. A sufficient number of replicas has to be
used in order to predict accurately a minimum energy path (MEP), for
most cases we repeated the calculations with a different number of
replicas until convergence of the activation energy and main features
of the MEP were observed. The total number of images actually used in
each case is reported where relevant in the following sections.

\section{Theoretical Results}\label{results}

\subsection{Bulk Mg, Ti and Ni, and the Mg(0001) surface}\label{bulk}

Magnesium bulk crystal at ambient conditions has the hexagonal closed
packed (hcp) structure.  Several preliminary tests were first carried
out using the PBE version of the PAW potential of magnesium, which
only includes the 3$s^2$ electrons in valence and has a core radius of
1.1~\AA. These included: the energy dependence on the $c/a$ ratio for
different {\bf k}-points meshes, from a minimum of 56 to a maximum of
880 {\bf k}-points in the irreducible wedge of the Brillouin zone
(IBZ); different values of $n$ for the MP smearing functions and
different smearing widths; and different plane-wave cutoffs.  To
calculate the bulk structural properties of Mg, energy versus volume
curves were fitted to a Birch-Murnaghan equation of
state~\cite{murna}.  We found that with a 18x18x12 {\bf k}-point mesh 
(259 points in the IBZ), $n=1$ and a smearing width of 0.2 eV, the zero 
pressure equilibrium volume $V_0$ and bulk modulus $B_0$ of bulk Mg were 
converged to within 0.2\% and 0.3\% respectively. Similar convergence 
results were obtained with the standard LDA potentials, which also has 
only the 3$s^2$ electrons in valence and a core radius of 1.1~\AA.  
Results are summarised in Table~\ref{table1}, together with results 
from previous theoretical calculations. The well known trend of LDA 
to overestimate the bulk modulus and underestimate the lattice 
parameter~\cite{wacho} is apparent.

Finally, we have tested PBE and LDA PAW potentials with 2$p^6$ and
3$s^2$ electrons in valence. These potentials still have core radii 
of 1.1~\AA\, but require an higher energy cut-off value of 350 eV. 
The structural parameters obtained with these potentials are essentially 
identical to those obtained with the previous potentials. Therefore, 
in the rest of the work we only used the standard Mg potentials.

\begin{table*}
\caption{\label{table1}Bulk and surface properties of pure Mg.}
\begin{ruledtabular}
\begin{tabular}{llcccc}
 & a (\AA) & c/a & V$^{(cell/atom)}_0$ (\AA$^3$) & k$_0$ (GPa) & dk$_0$ \\
\hline
This work & 3.19(3.13)\footnotemark[1] & 1.621(1.621)\footnotemark[1] & 22.85(21.59)\footnotemark[1] & 
36.8(40.5)\footnotemark[1] & 3.9(4.3)\footnotemark[1]  \\
Other calculations & 3.19\footnotemark[2], 3.18(3.13)\footnotemark[3], & 1.615(1.616)\footnotemark[3], & 
22.97(21.66)\footnotemark[4] & 35.5(40.2)\footnotemark[3], & 4.0(4.1)\footnotemark[4] \\
& 3.20(3.14)\footnotemark[4] & 1.624(1.622)\footnotemark[4] & &  36.0(40.1)\footnotemark[4] &  \\ 
$[$Expt.$]$ & [3.21]\footnotemark[5] & [1.624]\footnotemark[5] & [23.24]\footnotemark[6] & 
[35.4]\footnotemark[5], [36.8 $\pm$ 3.0]\footnotemark[7] & [4.3 $\pm$ 0.4]\footnotemark[7] \\ 
& & & & &  \\ 
\mbox{E$^{coh}$(eV)} & -1.50\footnotemark[1], -1.50(-1.78)\footnotemark[3] & & & &  \\ 
$[$Expt.$]$  & -1.51\footnotemark[8] & & & &  \\ 
& & & & &  \\ 
E$^{surf}$(eV/atom) & \multicolumn{5}{l}{0.30\footnotemark[1], 0.30(0.35)\footnotemark[3], 
0.34\footnotemark[9], 0.32\footnotemark[10]}   \\
$[$Expt.$]$  & 0.28\footnotemark[11], 0.33\footnotemark[12] & & & & \\  
\footnotetext[1]{Reported values are those from PAW PBE(LDA) calculations which do not include 
room temperature thermal expansion.}
\footnotetext[2]{Ref.~\onlinecite{du05}.}
\footnotetext[3]{Ref.~\onlinecite{wacho} from DFT GGA(LDA) calculations.}
\footnotetext[4]{Ref.~\onlinecite{mehta06} from PAW GGA(LDA) calculations.}
\footnotetext[5]{Ref.~\onlinecite{ashmerm}.}
\footnotetext[6]{Ref.~\onlinecite{webofel}.}
\footnotetext[7]{Ref.~\onlinecite{errandonea03}.}
\footnotetext[8]{Ref.~\onlinecite{kittel}.}
\footnotetext[9]{Ref.~\onlinecite{wright94} from ab initio LDA calculations.}
\footnotetext[10]{Ref.~\onlinecite{ismail00} from ab initio LDA calculations.}
\footnotetext[11]{Ref.~\onlinecite{hayden81}.}
\footnotetext[12]{Ref.~\onlinecite{tyson77}.}
\end{tabular}
\end{ruledtabular}
\end{table*}

From our PBE calculations (see Table~\ref{table1}), we derive a
lattice constant of 3.19~\AA, in error of just 0.6\% with respect to
the experimental value~\cite{ashmerm}. The zero pressure bulk modulus
$B_0$ is 36.8 GPa, and the value of $c/a$ at the equilibrium volume is
1.621, both in very good agreement with the experimental values (we
note however that these calculations do not include room temperature
thermal expansion, which are present in the experimental data).

Titanium is also hcp crystal. We used the standard version of the PBE
PAW pseudopotential for Ti, which has a core radius of 1.2 \AA, and we
used a 18x18x12 {\bf k}-point grid.  The resulting values for the
structural parameters were $a_0 = 2.923$ \AA\, and $B_0 = 120$~GPa,
and the value of $c/a$ at the equilibrium volume was 1.583, in good
agreement with those previously found with theoretical and
experimental investigations (see Refs.~\onlinecite{vohspen01,zopmish03,jahnatek05} 
and references therein).

To study bulk Ni we also used the standard version of the PBE PAW
potential, which has a core radius of 1.1~\AA. Ni bulk has a face
centred cubic crystal structure, with a small magnetic moment of 0.61
$\mu_B$ under ambient conditions~\cite{kittel}, so we performed spin
polarised total energy calculations. The calculations were performed
with a 13x13x13 grid of {\bf k}-points. We found a lattice parameter
$a_0=3.524$~\AA\, and a bulk modulus $B_0=194$ GPa, which compare well
with the experimental data of 3.524~\AA\, and 186 GPa
respectively~\cite{kittel,handbook}.  The zero pressure magnetic
moment is 0.63 $\mu_B$, which is also close to the experimental value
of 0.61 $\mu_B$. The values we found for $a_0$, $B_0$ and $\mu_B$ are
in agreement with those from other GGA and PBE
calculations~\cite{kresshafn00,greemavri05,pozzo07}.
 
Surfaces have been modelled using periodic slabs, with several atomic
layers (from 3 to 13) and a large vacuum thickness (5-18~\AA), defined
as the distance between two opposite facing surfaces.  We used a
18x18x1 {\bf k}-point grid for the 1x1 surface primitive cell.  The
positions of the atoms in the three topmost layers were allowed to
relax, while the rest were kept at the bulk interatomic distances.
Good convergence in the calculated surface energies and relaxations of
the topmost atomic layers was achieved with five layer slabs
(corresponding to a slab thickness of about 13~\AA) and a vacuum
region thickness of about 10~\AA. We found that the topmost layer has
an inward relaxation of about 1.4 \%, in good agreement with the
inferred experimental zero temperature value of 1.7\%~\cite{davis92}.

We found that with 5 atomic layers the surface energy is converged to
within 2 meV to the value 0.30~eV/atom. This compares well with the
experimental findings which are in the range
0.28-0.33~eV/atom~\cite{hayden81,tyson77}.


\subsection{H$_2$ dissociation and H diffusion on the pure Mg and the Ti-doped Mg surfaces}\label{H2dissMg}

Hydrogen adsorption energies on the Mg(0001) surface were determined
at low coverage in four possible sites: top, bridge, hollow-fcc and
hollow-hcp (see Fig.~\ref{adssites}). These adsorption energies are 
defined as $E_{\rm ads}$(H)$=$ $E_{\rm slab}$(MgH)-[$E_{\rm slab}$(Mg) 
+ 1/2 $E$(H$_2$)], where $E_{\rm slab}$(MgH) is the energy of the slab 
with one H adsorbed on the surface, $E_{\rm slab}$(Mg) is the energy 
of the bare slab and $E$(H$_2$) the energy of the isolated hydrogen molecule,
calculated by placing the H$_2$ molecule in a large cubic box of sides
13.5~\AA.

\begin{table}
\caption{\label{table2a} Hydrogen adsorption energies (E$_{ads}$) in different adsorption 
sites on the pure Mg surface, for the 2x2 and the 3x3 surface unit cells.}
\begin{ruledtabular}
\begin{tabular}{lcc}
Ads. sites & E$_{ads}$ (eV) & E$_{ads}$ (eV) \\ 
 & (2x2) & (3x3) \\
\hline
Top &  0.75 & 0.74 \\
bridge & 0.12 & 0.13 \\
hollow (hcp) & -0.03 & -0.03 \\
hollow (fcc)  & -0.05 & -0.04 \\ 
\end{tabular}
\end{ruledtabular}
\end{table}

Calculations have been performed on 2x2 (corresponding to 0.25 ML
coverage, ML=monolayer) and checked against results obtained from 3x3
(corresponding to 0.11 ML coverage) surface unit cells, with
differences between the two sets of calculations of less than 0.01~eV,
thus implying that the results effectively correspond to those for an
isolated H$_2$ molecule. The two sets of calculations have been
performed with equivalent grids of {\bf k}-points, 9x9x1 and 6x6x1 for
the 2x2 and the 3x3 surface unit cells respectively.

The values for the adsorption energies of atomic hydrogen in different
adsorption sites on the pure Mg surface are reported in
Table~\ref{table2a}.  These compare well with previous theoretical
results~\cite{bird93}.  It is clear that there is a strong preference
for the hollow sites, with a small preference for the fcc hollow site.

\begin{figure}
\rotatebox{270}{\scalebox{0.7}[0.8]{\includegraphics[width=\columnwidth]{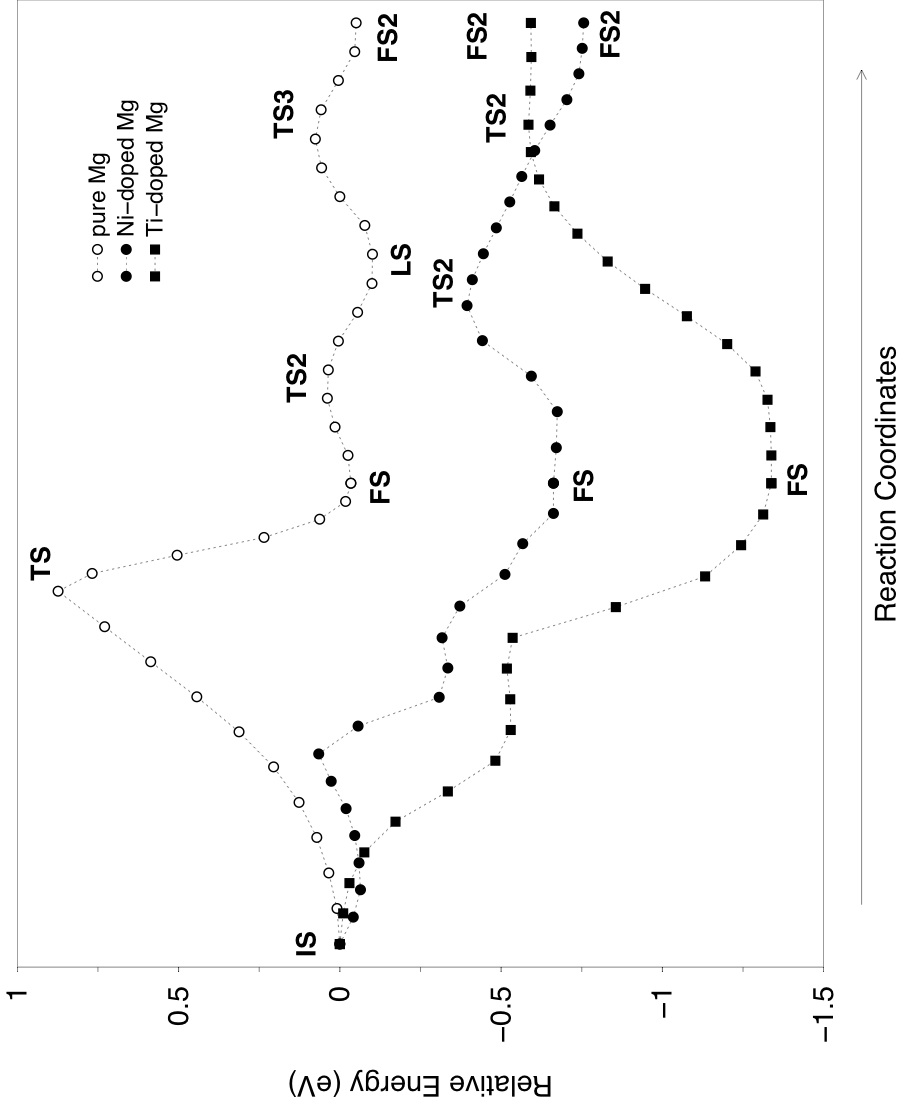}}}
\caption{\label{actbar} Minimum Energy Path for H$_2$ dissociation and diffusion on a pure 
Mg(0001), Ni-doped Mg(0001) and Ti-doped Mg(0001) surface.}
\end{figure}

We performed NEB calculations for H$_2$ dissociation over two possible
sites (bridge and top). These have been accompanied by careful tests
on supercell size, number of layers in the slab and number of replicas
in the NEB calculation to obtain the minimum energy path and the
activation barrier.  We found that a 2x2 supercell, 5 layers and 5
replicas are enough to obtain activation energies converged to within
0.02 eV. The first MEP is rather featureless (IS $\rightarrow$ TS
$\rightarrow$ FS), and it is well approximated also by 5 replicas,
although in Fig.~\ref{actbar} we report the MEP obtained with 17
replicas.
 
Of the two sites investigated, we found that H$_2$ prefers to
dissociate over a bridge site (see Fig.~\ref{MgH2diss}) with an
activation energy of about 0.87 eV (about 0.6 eV lower than that
obtained for dissociation over a top site), in agreement with previous
DFT calculations~\cite{norskhoum81,bird93,vegge04,du05} and
experimental findings~\cite{sprunplu91}, as reported in
Table~\ref{table2b}.


\begin{table}
\caption{\label{table2b} Activation energy (E$_a$) for hydrogen dissociation on the pure Mg, 
Ni-doped and Ti-doped Mg surfaces.}
\begin{ruledtabular}
\begin{tabular}{lc}
E$_a$ (pure Mg)  & 0.87\footnotemark[1], 0.4\footnotemark[2]$^,$\footnotemark[3], 
0.5\footnotemark[4]$^,$\footnotemark[5], 1.15\footnotemark[6], 1.05\footnotemark[7],
0.95\footnotemark[8] \\
$[$Expt.$]$ & 1.0\footnotemark[9] \\ \\
E$_a$ (Ni-doped Mg) & 0.06\footnotemark[1]\\ 
E$_a$ (Ti-doped Mg) & null\footnotemark[1], negligible\footnotemark[7]\\
\end{tabular}
\end{ruledtabular}
\footnotetext[1]{This work.}
\footnotetext[2]{Ref.~\onlinecite{hjelmberg79} for a jellium system.}
\footnotetext[3]{Ref.~\onlinecite{bird93}, from DFT LDA calculations ans PES. 
This lower value as compared to other calculations is explained as due to the well known 
LDA overbinding.}
\footnotetext[4]{Ref.~\onlinecite{johansson81} for a jellium system.}
\footnotetext[5]{Ref.~\onlinecite{norskhoum81} for a jellium system and PES.}
\footnotetext[6]{Ref.~\onlinecite{vegge04} from DFT RPBE.}
\footnotetext[7]{Ref.~\onlinecite{du05}, from DFT PAW RPBE calculations (see also discussion 
in main text).}
\footnotetext[8]{Ref.~\onlinecite{arboleda04} from PES calculations.}
\footnotetext[9]{Ref.~\onlinecite{sprunplu91}.}
\end{table}

The small difference between our findings and those of
Vegge~\cite{vegge05} and Du et al.~\cite{du05} are due to their use of
RPBE instead of PBE, and different {\bf k}-point meshes.

We then performed a second NEB calculation to obtain the MEP for the
diffusion of one of the H atoms on the surface from one fcc to a
second fcc site (FS $\rightarrow$ TS2 $\rightarrow$ LS $\rightarrow$
TS3 $\rightarrow$ FS2; see Fig.~\ref{MgH2diff}). This MEP (calculated 
with 17 replicas) is also displayed in Fig.~\ref{actbar}, as a continuation 
of the dissociation MEP, and shows that the highest energy barrier for 
surface diffusion is only $\sim 0.18$~eV, which agrees very well with the 
calculations of Du et al.~\cite{du07}. This low energy barrier clearly indicate
fast diffusion even at room temperature.

Before repeating the calculations on the Ti-doped surface, we tested
all four possible sites for H adsorption after dissociation (see
Fig.~\ref{holsites}), and we found that atomic hydrogen prefers to
adsorb into two of the possible three hollow-fcc sites around the Ti
atom.
The dissociation activation barrier was calculated using 9 and 17
replicas, with 9 being enough to display the main features of the MEP
(IS $\rightarrow$ FS; see Fig.~\ref{MgTiH2diss}), although in Fig.~\ref{actbar} 
we display the results obtained with 17 replicas.  Our findings are very similar to
the previous results of Du et al.~\cite{du05,du07}, i.e. there is no
barrier for hydrogen dissociation on a Ti-doped Mg surface, and a
barrier of almost 0.8 eV for diffusing away from the Ti sites (FS
$\rightarrow$ TS2 $\rightarrow$ FS2; see Fig.~\ref{MgTiH2diff}), which 
therefore becomes the rate limiting step in the reaction~\cite{du07}.




\subsection{H$_2$ dissociation and diffusion on a Ni-doped Mg surface}\label{H2dissMgNi}

Having benchmarked our calculations on the pure Mg and the Ti-doped Mg
surfaces, we now come to the main purpose of the paper, which is to
study the effect of Ni doping of the Mg(0001) surface on the
activation barriers for H$_2$ dissociation and diffusion on the
surface.

On the Mg(0001) surface, we found that Ni is non-magnetic, so all
calculations have been performed without including spin-polarisation.

\begin{figure}
\rotatebox{0}{\scalebox{0.6}[0.45]{\includegraphics[width=\columnwidth]{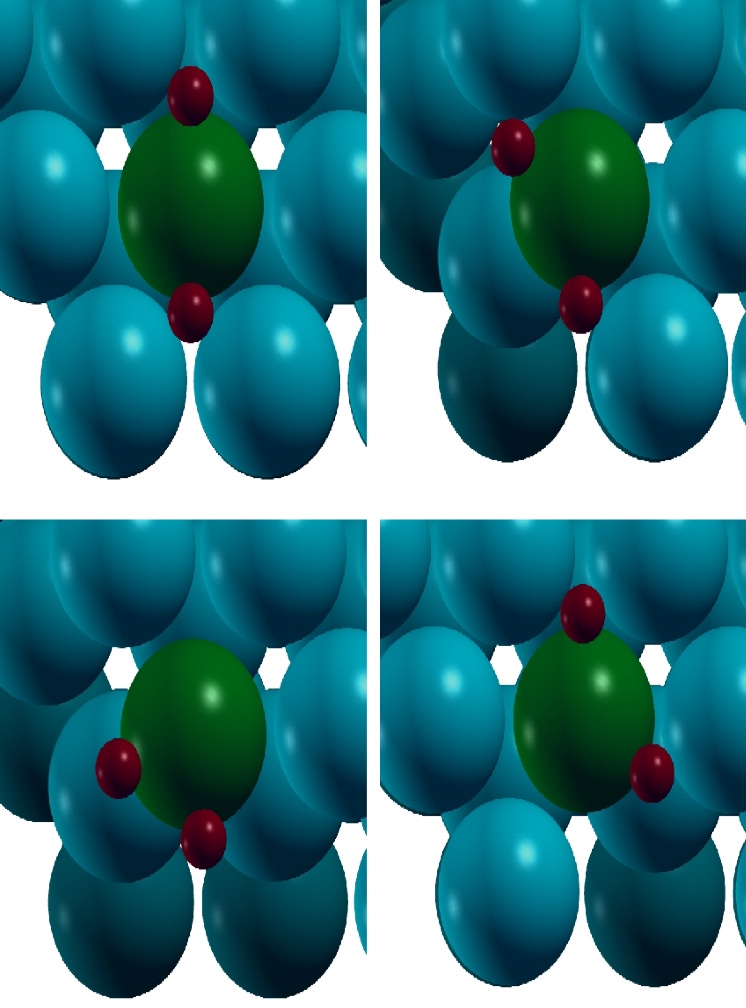}}}
\caption{\label{holsites} (Colour) Possible final state adsorption sites for H$_2$ 
(dark red) dissociation over the metal-doped (Ti/Ni) (dark green) Mg surface 
(light blue). In the case of the Ni doped surface the bottom-left site was not 
a stable configuration.}
\end{figure}

\begin{figure}
\rotatebox{-90}{\scalebox{0.8}[0.9]{\includegraphics[width=\columnwidth]{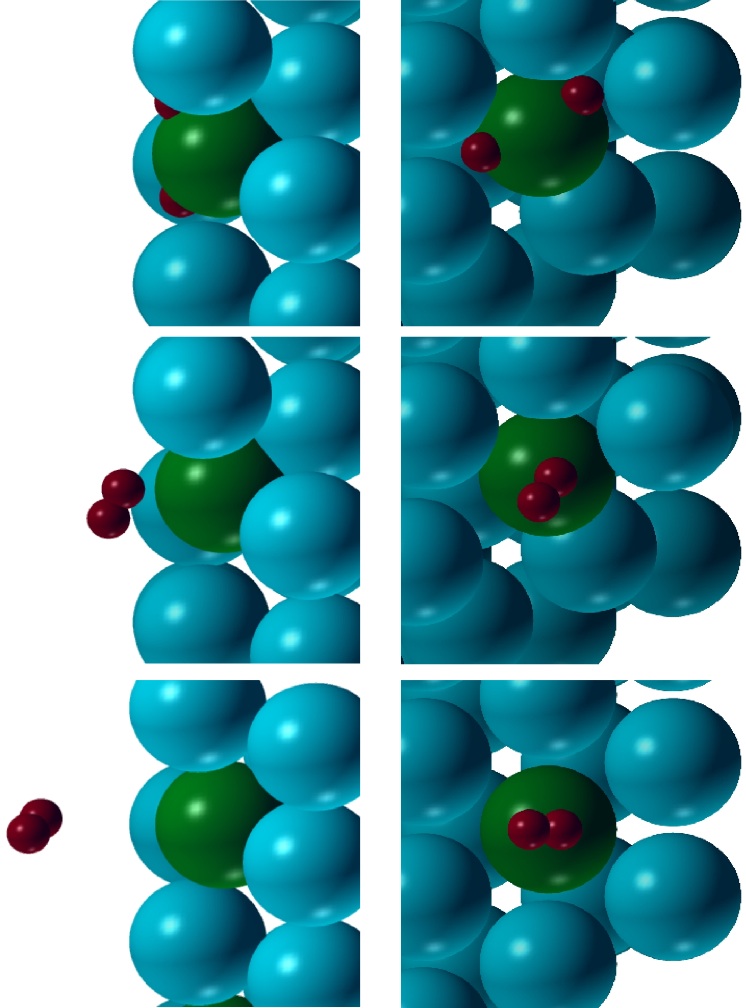}}}
\caption{\label{MgNiH2diss} (Colour) Same as Fig.~\ref{MgH2diss} but for H$_2$ dissociating
over the Ni-doped Mg surface. The Mg, Ni and H atoms are represented respectively
by light blue, dark green and dark red colours.}
\end{figure}

After dissociating on top of a Ni atom, the two H atoms can adsorb
into four different hollow sites, as shown in Fig.~\ref{holsites}. The
most stable final state is found to be the one where the H atoms
adsorb into two nearby hollow-hcp sites (see Fig.~\ref{holsites},
bottom-right corner. We also found that the configuration on the
bottom-left corner was unstable, with the hydrogen atoms repelled by
the Nickel atom and squeezed between nearby Mg
atoms). Figure~\ref{MgNiH2diss} shows the dissociation of the hydrogen
molecule over the Ni atom as viewed from side (top panel figures) and
top (bottom panel figures) positions respectively at the IS, TS and
FS.  Note that on the Ni-doped Mg surface the molecule at the TS is
much higher than on the pure Mg (0001) surface (see respectively
Figs.~\ref{MgNiH2diss} and \ref{MgH2diss}), being at $\sim 2$~\AA and
$\sim 1$~\AA~over the two surfaces respectively.

NEB calculations were run with different numbers of replicas, and we
found that 9 replicas are enough for a precise calculation of the
energy barrier to within 1 meV. The resulting activation barrier for
H$_2$ dissociation on a Ni-incorporated Mg surface is only 0.06 eV,
against 0.87 eV found for the pure Mg surface. In Fig.~\ref{actbar} we
display the MEP obtained from a calculation with 17 replicas (IS
$\rightarrow$ TS $\rightarrow$ FS).

The NEB diffusion calculation was also performed with 17 replicas
(FS $\rightarrow$ TS2 $\rightarrow$ FS2; see Fig.~\ref{MgNiH2diff}) and shows 
an energy barrier of only 0.27 eV, which is only slighlty higher than the 
diffusion barrier on the pure Mg surface, and would also allow fast diffusion 
even at room temperature.  We note that this barrier is similar to the one 
found on the Pd-doped surface by Du at al.~\cite{du07}, although in that case 
the rate limiting step is the dissociation of the H$_2$ molecule with an 
energy barrier of 0.305~eV.
 
This suggests that Ni should be an even better catalyst than Pd for
the hydrogenation of Mg.


 
As a final note we would like to point out an interesting analogy with 
H$_2$ dissociation on pure transition metal surfaces. In particular, 
on the pure Ni(111) surface Kresse~\cite{kresse00} calculated an energy
barrier of only 0.015 eV using DFT PAW GGA. This is similar to our
value of 0.06 eV on the Ni-doped Mg surface, however, we note that
when the same 4x4x1 {\bf k}-point sampling grid is used we find an
energy barrier of 0.014 eV on the Ni-doped Mg surface, which is
therefore very close to the value found by
Kresse~\cite{kresse00}. These calculations are also consistent with
potential energy calculations of Arboleda et al.~\cite{arboleda04},
also performed with a 4x4x1 {\bf k}-point grid. The small barrier
for hydrogen dissociation on Ni(111) is also confirmed by
experiments~\cite{rendulic89}).

Analogously, the behaviour of H$_2$ dissociation over the Ti-doped Mg
surface appears to be similar to that obtained on the pure Ti(111)
surface: the null activation barrier we find with a smaller 4x4x1 grid
compares with theoretical results found over a Ti (0001) surface with
an analogous grid~\cite{nobuhara04,arboleda04}. In other words, this
seems to suggest that the value of the activation barrier for hydrogen
dissociation over a transition-metal doped Mg surface is similar to
the activation barrier for H$_2$ dissociation over the corresponding
pure transition-metal surface.

\begin{table*}
\caption{\label{table3} The d-band center position with respect to the
Fermi energy ($E_d - E_f$), H $s$ peak shift between the initial and
transition state (H$_s$$^{TS-IS}$), activation barrier (E$_{a}$) and
energy difference between the final and initial state (E$^{FS-IS}$)
for hydrogen dissociation on the pure Mg surface as opposed to the
Ni/Ti-doped Mg surfaces.}
\begin{ruledtabular}
\begin{tabular}{lcccc}
Surface & $E_{db} - E_F$ (eV) & H$_s$$^{TS-IS}$ (eV) & E$_{a}$ (eV) & 
E$^{FS-IS}$ (eV) \\

\hline
Mg pure & -- & -1.43 & 0.87 & -0.04 \\
Ni-doped Mg & -0.79 & -0.77 & 0.06 & -0.66 \\
Ti-doped Mg & +1.08 & -- & null & -1.34 \\
\end{tabular}
\end{ruledtabular}
\end{table*}



\subsection{Electronic structure}\label{dos}

To study the electronic properties of the system, we projected the
electronic density of states onto spherical harmonics functions of type
$s$, $p$ and $d$, centred on Mg, Ni, Ti and H atoms.  It is well known that
the catalytic reactivity of a surface is correlated to the position of
the $d$-band (i.e., in this case the projection of the electronic
density of states onto $d$ type spherical harmonics) with respect to the
Fermi energy $E_f$. In particular, it was shown by Hammer and
Norskov~\cite{hamnor95}(see also~\cite{mavrikakis98}) that a convenient
parameter to monitor the catalytic reactivity is the first energy
moment of the $d$-band, or $d$-band centre, defined as $E_d =
\int_{-\infty}^{E_0} dE (E-E_f) p_d(E)$, where $p_d(E)$ is the density
of states projected onto spherical harmonic of type $d$ centred on some
specified atom, and $E_0$ is some cutoff energy which we chose to be
at 7 eV above the Fermi energy.  Then, if the centre of the band is
close to $E_f$ it follows that there are many $d$ electrons available
for donation, as well as a significant number of empty $d$-levels
available for back-donation, and the results of this is that the
system is very reactive. The $d$ electrons of Ni on Mg(0001) form a
band which is relatively close to the Fermi energy and for this reason
the system is very reactive. By contrast, Mg has no $d$ electrons
(although in the solid state a projection onto $d$ type spherical
harmonics is not zero), and therefore its reactivity is much reduced
by comparison.

Ni is a late transition metal with almost all the $d$ orbitals filled
with electrons, by contrast Ti only has 2 electrons in the $d$
orbitals. It is therefore clear that the position of the $d$-band
centre will be much higher in Ti than in Ni, which explains the higher
reactivity of Ti.

Our calculated values for $E_d$ on the Ni/Ti-doped Mg surfaces are
-0.79 eV and +1.08 eV for Ni and Ti respectively (see Table~\ref{table3}).

Using DFT RPBE, Vegge et al.~\cite{vegge05} calculated the $d$-band center 
position for MgTM (TM $=$ transition metal) alloys, allowing an expansion
of the alloy lattice to accomodate the hydrogen atoms. They found -0.82 and 
+0.48 for TM=Ni,Ti respectively. Although the value we find for Ti is much larger, 
the same trend is observed in the case of our Ni/Ti-doped Mg surfaces.

\begin{figure}
\rotatebox{0}{\scalebox{1.2}[0.8]{\includegraphics[width=\columnwidth]{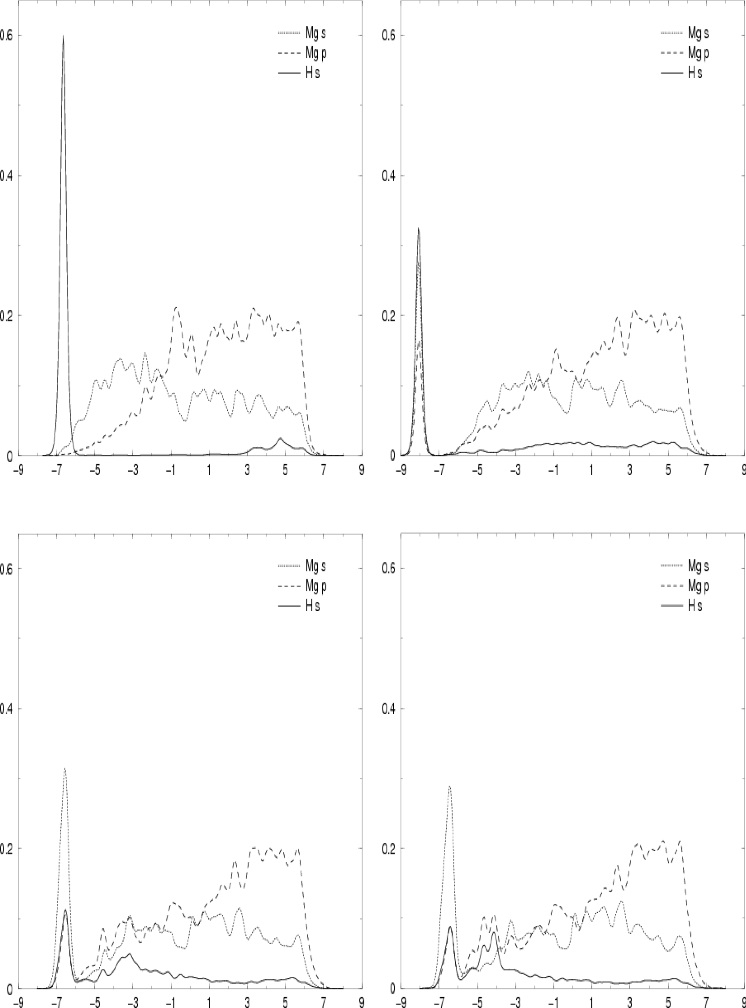}}}
\caption{\label{dosMg} Projected density of states for H$_2$
dissociating over a pure Mg surface as a function of the energy
relative to the Fermi level, respectively for the initial state (IS,
top-left corner), transition state (TS; top-right corner), transition
state plus one further step along the MEP (TS+1; bottom-left corner)
and final state (FS; bottom-right corner).}
\end{figure}

\begin{figure}
\rotatebox{0}{\scalebox{1.2}[0.8]{\includegraphics[width=\columnwidth]{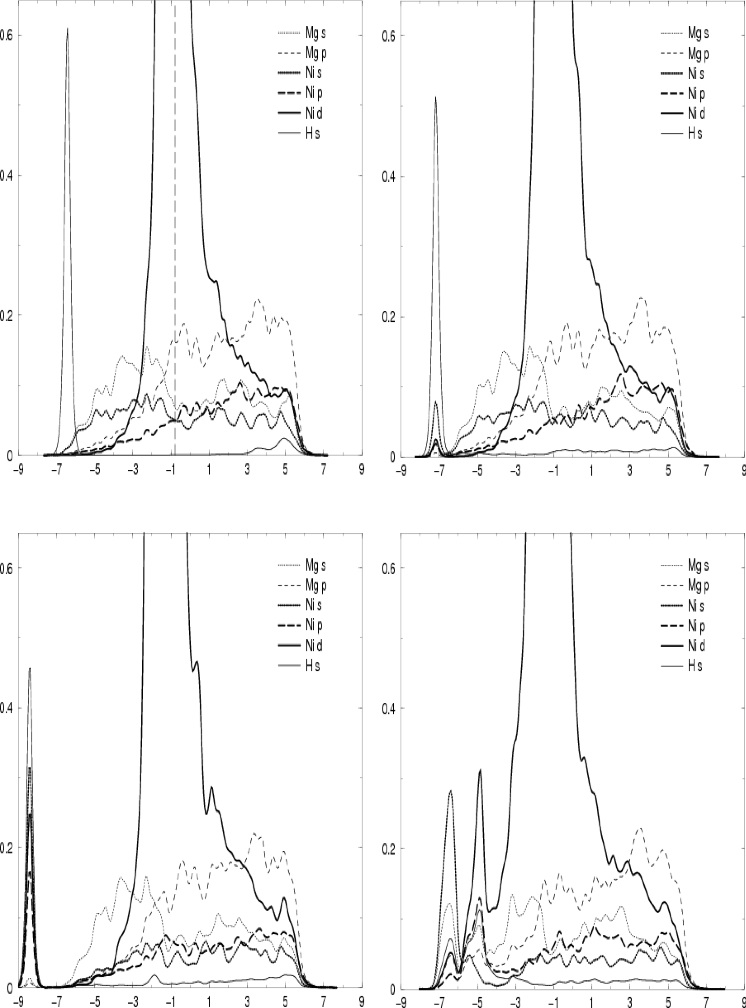}}}
\caption{\label{dosMgNi} As in Fig.~\ref{dosMg} but for the Ni-doped Mg surface.
The dashed vertical line shows the position of the d-band centre.}
\end{figure}

\begin{figure}
\rotatebox{0}{\scalebox{1.2}[0.4]{\includegraphics[width=\columnwidth]{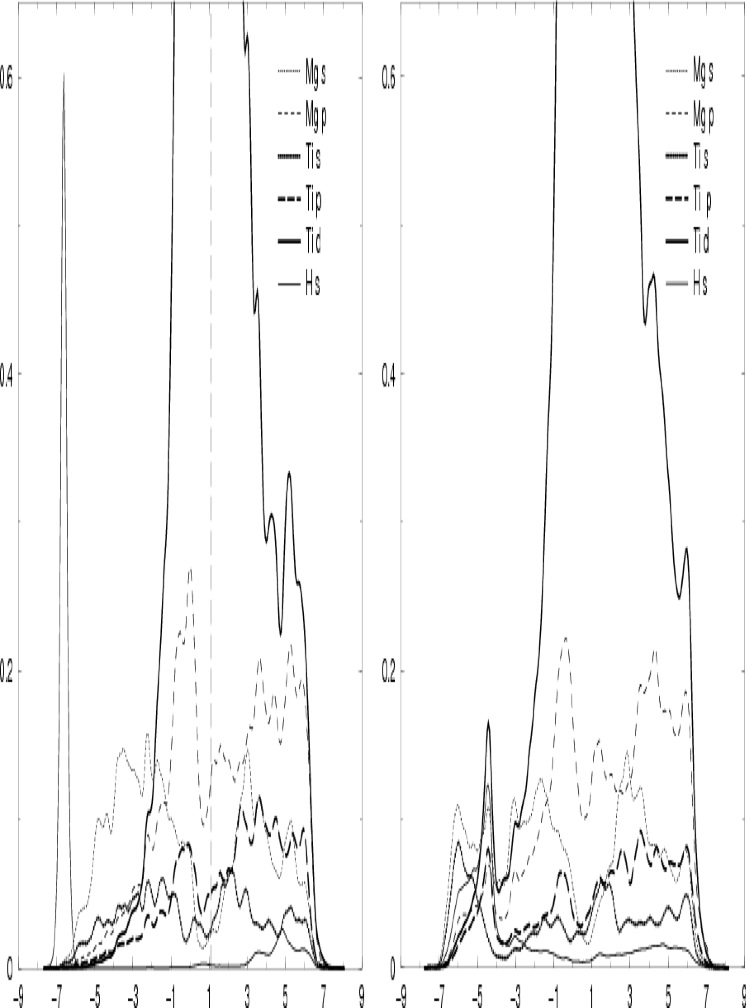}}}
\caption{\label{dosMgTi} As in Fig.~\ref{dosMg} but for the Ti-doped Mg surface.
Note that there is no barrier for hydrogen dissociation for this surface, therefore
the dos are those for IS and FS only. The dashed vertical line shows the position 
of the d-band centre.}
\end{figure}

Figures~\ref{dosMg}-\ref{dosMgTi} show the projected density of states
(PDOS) found for the pure Mg, Ni-doped and Ti-doped Mg surfaces
respectively. The PDOS are given for a number of configurations along
the MEP: the initial state (IS), the transition state (TS), the
replica just after the transition state (TS+1) and the final state
(FS).  For the Ti-doped Mg surface, the PDOS are given for IS and FS
only since there is no transition state in the dissociation
process. For somplicity of notation, we call here the transition state
and the final state simply TS and FS, as we only refer to the part of
the MEP which deals with the dissociation of the H$_2$ molecule.

In the IS the hydrogen molecule is still far from the metal surface
and there is no overlap between the H$_2$ molecular orbitals and the
orbitals of the metal surface.  At the transition state, instead, when
gaseous hydrogen has started dissociating over the surface, there is
clear interaction between the H $s$ orbital and the Mg $s$ and $p$
orbitals on the pure Mg surface (see Fig.~\ref{dosMg}, top-right
corner). On the Ni-doped surface the overlap appears to be non-zero
with all the Ni $s$, $p$ and $d$ orbitals (see Fig.~\ref{dosMgNi},
top-right corner).  In the final state it is evident that the
magnitude of the Mg $p$ electron peaks below the Fermi level are
increased in the Ni/Ti-doped surfaces (see respectively
Fig.~\ref{dosMgNi}, bottom-right corner, and Fig.~\ref{dosMgTi},
right) with respect to the pure Mg surface (see Fig.~\ref{dosMg},
bottom-right corner).

Interestingly, we note that there appears to be a clear negative shift
of the position of the hydrogen $s$ orbital in going from the initial
state to the transition state, which is more pronounced for the pure
Mg surface as opposed to the Ni-doped Mg surface.

Besides the $d$-band center positions, in Table~\ref{table3} we also
report the corresponding activation barriers ($E_{a}$), the energy
difference between the initial and final states ($E^{FS-IS}$) for
hydrogen dissociation, and the H $s$ peak shift between the initial
and transition states (H$_s$$^{TS-IS}$). The correlation between the
position of the $d$-band and the height of the activation barrier is
evident, as well as the correlation with $E^{FS-IS}$, i.e., the 
$d$-band center position is smaller for larger values of the former
and smaller values of the latter.

Furthermore, from the results obtained for the pure Mg and Ni-doped Mg
surfaces, another interesting correlation emerges. In fact, as shown
in Table~\ref{table3}, it appears that H$_s$$^{TS-IS}$ correlates with
both $E_{a}$ and $E^{FS-IS}$, i.e., it is smaller for smaller values
of the former and for larger values of the latter, following a
reversed trend with respect to that noticed for the $d$-band center
position.  In other words, this means that the shift of the hydrogen
$s$ orbital between initial state and transition state is larger when
the bond between the dopant and H atoms is weaker.

\subsection{Charge distribution}

To conclude our analysis we decided to have a look at the charge
distributions in the systems as the dissociation processes take place
on the pure Mg and metal-doped Mg surfaces.

\begin{figure}
\rotatebox{0}{\scalebox{1.2}[0.4]{\includegraphics[width=\columnwidth]{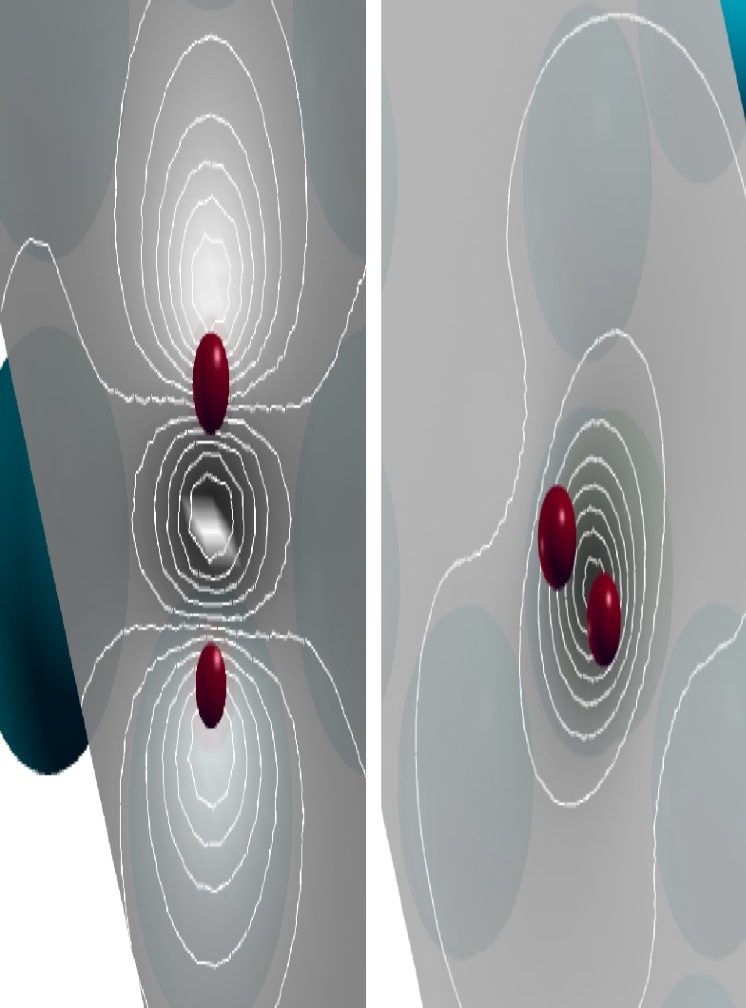}}}
\caption{\label{charge} (Colour) Charge distribution during H$_2$ (dark red) 
dissociation at the TS of the MEP respectively on the pure Mg (left) and 
Ni-doped Mg (right) surfaces (see text for details). White shows positive 
charge and black negative charge. Isolines are also shown in white.} 
\end{figure}

To do this, we calculated the total charge at each step of the MEP,
and for convenience, we subtracted the charge densities obtained from
calculations which included only the substrate and only the H$_2$
fragments respectively, with the atoms in exactly the same positions.
This charge difference obviously integrates to zero, and has the
advantage of showing point by point where the charge is being
transferred to.  The analysis reveals some interesting effects. In
particular, on the pure Mg(0001) surface we find that at the
transition state there is a significant charge transfer from the Mg
substrate to the H atoms (see Fig.~\ref{charge} - left).  This extra
charge fills the H antibonding orbitals which eventually leads to
dissociation, and builds up on the molecule because the Mg surface is
unwilling to accept back-donation of electrons from the H atoms.  The
Coulomb energy of this charge transfer is probably the main
contribution for the energy barrier.

By contrast, on the Ni doped surface there is almost no charge
transfer from the substrate to the molecule at the transition state
(see Fig.~\ref{charge} - right). This is because while some Ni $d$
charge fills up the H$_2$ antibonding orbitals, charge from the
molecular bonding orbital is back-donated to the empty $d$ states
available on the surface.  As a result, the energy barrier is reduced
to almost zero.

\section{Experimental}\label{exper}

\subsection{Sample preparation}

The three samples prepared were MgH$_2$, 2\%Ni/MgH$_2$ and
2\%Ti/MgH$_2$.  Three batches of 25 gram samples were prepared by ball
milling the different compositions for 2 hours under 4 bar of
hydrogen.  The MgH$_2$ used for all is Goldschmidt 98\% pure. The Ni
used (99.9\% pure) was from Alfa Aesar 0.8 - 0.03 $\mu$m diameter as
was the Ti used ($<$20 $\mu$m and 93\% pure). 25 g of both
2\%Ni/MgH$_2$ and 2\%Ti/MgH$_2$ were mixed in a tubular mixer before
milling for one hour. Samples were then milled using the Fritsch
planetary ball mill pulverisette 5.  The milling pots have a special
stainless steel jacket with an o-ring fitted on the top seal, this can
allow a gas atmosphere through a feeding valve, to be used during the
milling process of up to 5 bar.  25 grams of sample were milled using
agate pots (around 300cc volume) and 15 balls of the same
material. The milling process was 2 hours using 350 rpm in a 15 minute
mill 10 minute pause sequence.

\subsection{Sample testing}

The rig used for testing the sample has a 10 cc reactor pot containing
1 gram of sample. The main lines of the rig are a hydrogen line
regulated to a 7 bar gauge, an Argon line and a vacuum line. The inert
gas line and a vacuum line are used for purging the system. The
reactor is connected to an inlet flow controller, a pressure
transducer to read the internal pressure and a mass flow meter
outlet. A thermocouple in close contact with the powder load reads the
sample temperature. A heating jacket cartridge is attached to the
reactor allowing the system to operate in isothermal conditions or be
temperature programmed from a control box which uses the sample
thermocouple as a reference value. An interface card records inlet
flow, outlet flow, temperature and pressure every second.

Volumetric hydrogenation and dehydrogenation cycles were possible to
monitor using this arrangement. Hydrogenations were performed at 300 C
using 25 cc min$^{-1}$ of hydrogen from a regulator set at 7 bar
gauge. During the hydrogenation the pressure increases to a point
where the sample starts absorbing and forms a plateau pressure, once
the sample is fully hydrogenated the reactor keeps building pressure
until seven bar are reached.  Dehydrogenations were recorded by
flowing 25cc/min of hydrogen through the reactor using an inlet flow
controller. The system is then open to vent for a chosen temperature
or heating slope and any hydrogen evolving from the sample is recorded
as an increase in the 25cc min flow by an outlet flowmeter.

\subsection{Experimental Results}

\begin{figure}
\rotatebox{0}{\scalebox{1.0}[1.0]{\includegraphics[width=\columnwidth]{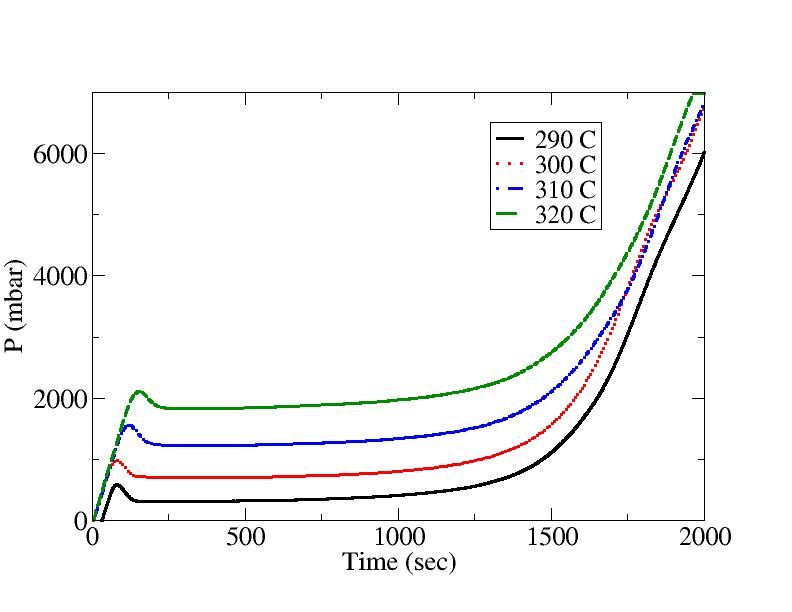}}}
\caption{\label{JM1} (Colour) Different temperature hydrogenation plots for 1 gram of 
2\%Ni/Mg using 25 cc/min of H$_2$.}
\end{figure}

\begin{figure}
\rotatebox{0}{\scalebox{1.0}[1.0]{\includegraphics[width=\columnwidth]{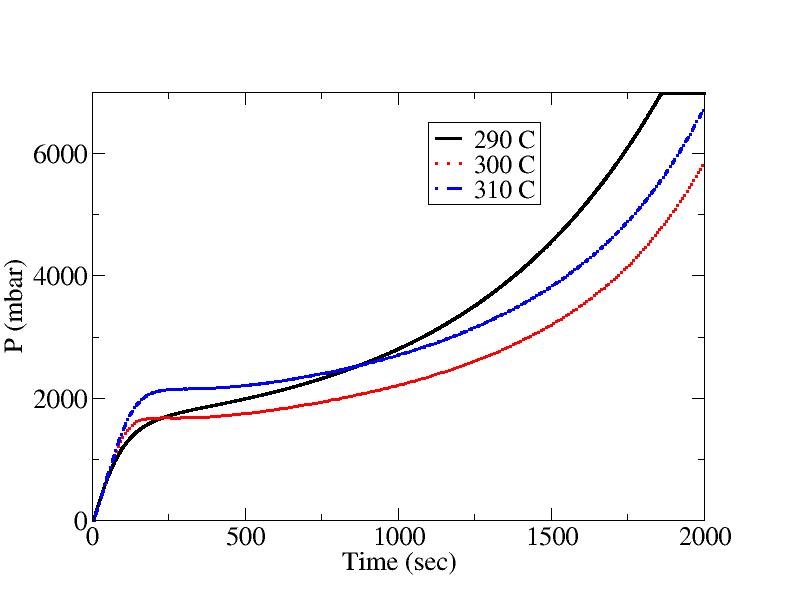}}}
\caption{\label{JM2} (Colour) Different temperature hydrogenation plots for 1 gram of 
2\%Ti/Mg using 25 cc/min of H$_2$.}
\end{figure}

The hydrogenation plots of the 2\% Ni/Mg sample at different
temperatures in the range between 290 and 320 C are shown in
Fig.~\ref{JM1}. The hydrogenation of the 2\%Ti/Mg sample in the
temperature range between 290 and 310 C is shown in
Fig.~\ref{JM2}. Both graphs of hydrogenation show P mbar gauge vs time
in seconds. The hydrogenation of pure Mg gave results close to the Ti
doped samples and is therefore not shown.

The catalytic activation of Mg by Ni during hydrogenation is
clear. The plateau pressures for each temperature hydrogenation are
lower for the 2\%Ni/Mg sample than those of Mg and
2\%Ti/Mg. The fact that at 290 C the hydrogenation curve for
2\%Ni/Mg is still lower than the hydrogenation in the same
conditions at 300 C confirms this. At 290 C both Mg and
2\%Ti/Mg showed a higher hydrogenation pressure than at 300 C
suggesting Ti catalysisof Mg hydrogenation is not evident.

\section{Conclusions}\label{conclusions}

We have presented a DFT study of hydrogen dissociation and diffusion
over Ni-doped and Ti-doped Mg(0001) surfaces, and compared these with
dissociation and diffusion on pure Mg(0001). Our results show that the
energy barrier for hydrogen dissociation is high on the pure Mg
surface (0.87 eV), and it is small (0.06 eV) or even null when Ni or
Ti are used as dopants respectively. We also found that although on
the pure Mg(0001) surface the binding energy of the H fragments is
nearly zero, on the Ni/Ti -doped surfaces this binding energy is
significant, being 0.66 and 1.34 eV respectively, with diffusion
energy barriers of $\sim 0.18$, $\sim 0.27$ and $\sim 0.8$eV on the
pure Mg, Ni-doped and Ti-doped surfaces respectively. 

Interestingly, the activation barriers for H$_2$ dissociation over the
Ni/Ti-doped Mg surface are similar to the values found on the
corresponding pure Ni/Ti surfaces
\cite{arboleda04,kresse00,nobuhara04}.

More insight in the behaviour of these systems can be gained by
inspecting the partial density of states and by looking at the
electronic charge density distributions.  In particular, the higher
reactivity of Ti with respect to Ni can be understood in terms of a
lower position of the $d$-band centre, which correlates with both the
height of the energy barriers for the dissociation of the H$_2$
molecule and with the binding energy of the H fragments when adsorbed
on the surface.

The charge density distributions on the different systems also shows
some interesting behaviour. In particular, we argued that the presence
of a barrier on the pure Mg(0001) surface may be understood in terms
of the build up of extra charge on the H$_2$ molecule as it
moves closer to the surface. This happens because the closed shell Mg
surface is unwilling to accept back-donation of charge from the H$_2$
molecule. One consequence of this is that the molecule needs to arrive
very close to the surface before starting to dissociate. By contrast, 
Ni and even more so Ti have many available empty $d$-states, and this 
avoids significant charge transfer from the substrate to the molecule.
In this case, the dissociation of the molecule begins much further away
from the surface.

The low dissociation barrier, coupled with the low diffusion barrier,
make Ni a very useful promoter for the hydrogenation of Mg. By
contrast, the high dissociation barrier on the pure Mg surface, and
the high diffusion barrier on the Ti-doped surface, are responsible
for the slow kinetics of hydrogenation on both systems.

Our experimental findings show faster hydrogenation for the 2\%Ni/Mg
sample with respect to the reference Mg or the 2\%Ti/Mg, in good
agreement with our theoretical results of a lower activation energy
for the dissociation-diffusion process in the 2\%Ni/Mg system. The
behaviour of Mg and 2\%Ti/Mg upon hydrogenation is found to be very
similar, again agreeing very well with the theoretical findings of
large and similar activation energies: a dissociation energy barrier
of 0.87 in the pure Mg system, and a diffusion energy barrier of 0.8
eV in the 2\%Ti/Mg system, making the dissociation-diffusion process
similarly difficult in both cases.

We deliberately chose to study Ni and Ti as dopants because they are
at the two ends of the first row of transition metals, and so their
behaviour may be expected to be representative of a range of
properties. In fact, we are now extending our investigations to other
transition metals, and we plan to report on these new results in the
near future. 

\newpage

\begin{acknowledgments}
This work was conducted as part of a EURYI scheme award as provided by
EPSRC (see www.esf.org/euryi). Allocation of computer time has been
provided by UCL research computing.
\end{acknowledgments}

\clearpage

\clearpage

\clearpage

\clearpage

\clearpage

\clearpage

\clearpage

\clearpage

\end{document}